\documentclass[a4paper]{article}

\usepackage{INTERSPEECH2020}
\usepackage{todo}
\usepackage{bm}

\title{Improving Unsupervised Sparsespeech Acoustic Models \\ with Categorical Reparameterization}

\name{Benjamin Milde, Chris Biemann}
\address{Language Technology Group, Dept. of Informatics, Universit\"at Hamburg}
\email{\{milde,biemann\}@informatik.uni-hamburg.de}

\begin{document}

\maketitle
\begin{abstract}

The Sparsespeech model is an unsupervised acoustic model that can generate discrete pseudo-labels for untranscribed speech. We extend the Sparsespeech model to allow for sampling over a random discrete variable, yielding pseudo-posteriorgrams. The degree of sparsity in this posteriorgram can be fully controlled after the model has been trained. We use the Gumbel-Softmax trick to approximately sample from a discrete distribution in the neural network and this allows us to train the network efficiently with standard backpropagation. The new and improved model is trained and evaluated on the Libri-Light corpus, a benchmark for ASR with limited or no supervision. The model is trained on 600h and 6000h of English read speech. We evaluate the improved model using the ABX error measure and a semi-supervised setting with 10h of transcribed speech. We observe a relative improvement of up to 31.4\% on ABX error rates across speakers on the test set with the improved Sparsespeech model on 600h of speech data and further improvements when we scale the model to 6000h.

\end{abstract}


\section{Introduction}

Transcribed and labeled speech data is usually needed to train supervised speech recognition systems, yet it is costly to obtain and transcribe. In contrast, unlabeled speech data can be obtained in vast quantities, even for languages for which much less resources are available as compared to e.g. English.

In recent years, unsupervised acoustic modelling has been gaining traction as viable models emerge to leverage and make use of a treasure trove of unlabelled speech data. The task of acoustic unit discovery has gained significant popularity in unsupervised or zero resource speech processing \cite{dunbar2017zero}. Unsupervised unit discovery in isolation can provide insights into datasets, phoneme modelling choices and ultimately provide representations that enables working with raw speech when transcriptions are completely absent.  

However, using what unsupervised acoustic models learn and transferring that knowledge in semi-supervised and transfer learning settings is of considerable practical interest. These learning settings hold the promise to boost performance of supervised systems, especially in low-resource settings. In this work, we extend and evaluate an unsupervised acoustic model originally proposed for acoustic unit discovery also in a semi-supervised setting. A large amount of untranscribed speech data (600h-6000h) and only a small amount (10h) of transcribed speech training data is available in this learning setting.

Using the masking technique for unsupervised modelling and then fine-tuning is reminiscent of transformer models such as BERT \cite{devlinetal2019bert}, that are currently very popular on text data. As speech is continuous, using the idea of masking becomes a bit more difficult to transfer directly. In the following we present our approach that is based on a memory component addressed by Gumbel-Softmax, as part of a larger recurrent encoder/decoder network. We then use the masking technique on the internal representation, i.e. the pseudo-posteriograms used for reconstruction and generated at each time step can be randomly masked.

\section{Related work}

The ZeroSpeech challenges \cite{versteegh2015zero, dunbar2017zero, dunbar2019zero} target speech processing in a zero resource setting, i.e. models that learn from raw speech without transcription. The challenges have also established the use of the ABX discriminability \cite{schatz2013evaluating, schatz2016abx} to intrinsically evaluate how well semantically relevant sounds are mapped by a discovered representation in the acoustic unit discovery task. 

Models trained on untranscribed speech are recently becoming relevant since they can also boost performance in supervised systems:

Schneider et al. \cite{schneider2019wav2vec} showed with Wav2vec that pre-training a model on raw speech with a similar binary contrastive loss as Word2Vec \cite{mikolov2013distributed} can be effective to improve supervised end-to-end acoustic models. The discrete variant vq-wav2vec \cite{baevski2019vq, baevski2020effectiveness} of this model with vector quantization into several thousand of units has also been successfully used to pretrain BERT \cite{devlinetal2019bert} on a sequence masking task, followed by using BERT representations in a wav2letter \cite{collobert2016wav2letter} acoustic model for speech recognition.

Vector quantized variational autoencoders \cite{van2017neural} can also be used to learn discrete representations of speech, as demonstrated by the end-to-end system involving attention based ASR and TTS \cite{tjandra2019vqvae} to encode and decode. Wang et al. proposed input masking \cite{wang2020unsupervised} in recurrent auto-encoders. 

Contrastive Predictive Coding (CPC) \cite{oord2018representation} is a representation learning model trained by predicting future hidden states, which can be applied to raw speech in the time domain.  In \cite{kahn2019libri}, Kahn et al. created a benchmark for ASR with no or limited supervision, based on English audio books and Librispeech data, also providing results for using CPC. We use the Libri-Light corpus for training and evaluating our models, as it provides a good benchmark for unsupervised acoustic models that can scale well to large amounts of (untranscribed) speech data.

\section{Sparsespeech model}

In \cite{milde2019sparsespeech}, we previously proposed an approach to train unsupervised bi-directional recurrent neural network (RNN) acoustic models that learn discrete representations, with a memory-augmented auto encoder. The Sparsespeech model also uses sequence masking (sequence dropout) on a quasi-symbolic representation that the network generates. The model consists of an encoder that generates the quasi-sparse representation of speech and a decoder that reconstructs the input features from embeddings of a memory component addressed with this quasi-sparse representation. Encoder and decoder are each a bi-directional stacked Long Short-Term Memory (LSTM). A continuous context vector is also an additional input to the decoder, with the idea to capture and entangle variability of utterance global factors such as speaker identity or the environment. It can be an explicit context representation \cite{milde2018unspeech} or speaker vector \cite{dehak2011front}; in this paper we use an implicit context vector that is the mean of all encoder states, as also evaluated in \cite{milde2019sparsespeech}. This also has the advantage that no separate model needs to be trained. When Sparsespeech representations are generated, we use output of the encoder's softmax. A sparsity constraint and diversity constraint on the encoders softmax output values $\sigma$ is used in the original model to train the model on continuous approximations of one-hot vectors:

\begin{equation}
\textnormal{Sparsity-} L = 1 - \sup_{n} \sigma_i
\end{equation}

\begin{equation}
\textnormal{Diversity-}L = \frac{1}{m} \sum_{j=1}^{m} D_{KL}(\sigma_j || U) 
\end{equation}

where $m$ are time steps of an utterance with $n$ softmax outputs per timestep using Kullback-Leibler (KL) divergence \cite{kullback1951information} and $U(x)=\frac{1}{n}$. A sparsity weight is multiplied with Sparsity-L and a diversity weight is multiplied with Diversity-L, these terms are then added to the loss function.

One drawback of this sparsity constraint is that it cannot be changed at generation time, as it is a hyper-parameter at training time. In this paper, we extend Sparsespeech to model symbolic self-labeling as an (approximated) discrete distribution, introducing an additional parameter that can be used to control the sparseness of the pseudo-posteriorgram representations that our model generates after training.

\section{Categorical reparameterization}

Discrete variables are difficult to train directly in a neural network, as the backpropagation algorithm cannot by applied to a non-differentable layer.

We use categorical reparameterization \cite{jang2017categorical} by Gumbel-Softmax \cite{gumbel1948statistical} to implement approximate discrete inference within the network while training it. This uses the softmax function as a differentiable approximation to argmax as follows: 

We sample a noise vector $\bm{g}=g_1 \dots g_k$ from a Gumbel distribution with a uniform random sampler $U$:

\begin{equation}
\bm{g} = -log(-log(U(0,1)))
\end{equation}

Where $k$ is the number of elements in the softmax. We then compute the Gumbel-Softmax as:

\begin{equation}
softmax(\frac{logits+\omega\cdot\bm{g}}{\tau} )
\end{equation}

Where $\omega$ is a noise weight parameter. We set $\omega =1$ while training the network and $\omega=0$ after the training is completed to disable the Gumbel noise.

\begin{figure}[t]
  \centering
  \includegraphics[width=\linewidth]{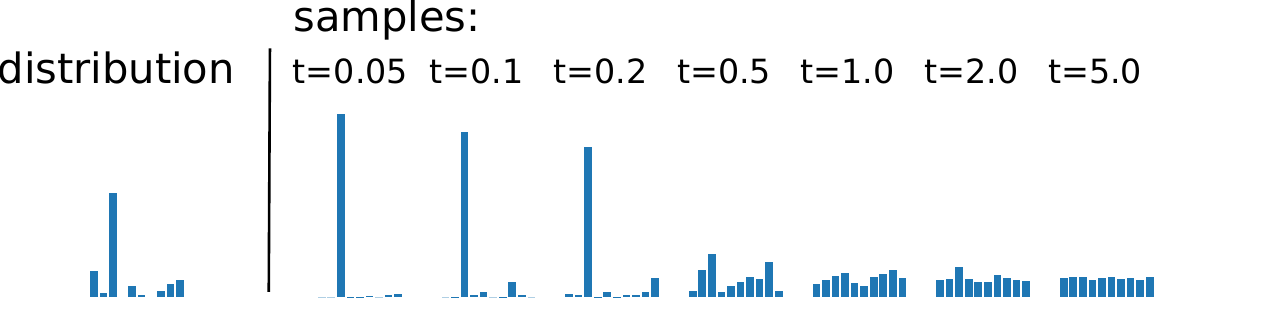}
  \caption{Drawing samples with different temperatures with the Gumbel-Softmax from a discrete distribution.}
  \label{fig:gumbel_ex}
\end{figure}

\begin{figure}[t]
  \centering
  \includegraphics[width=\linewidth]{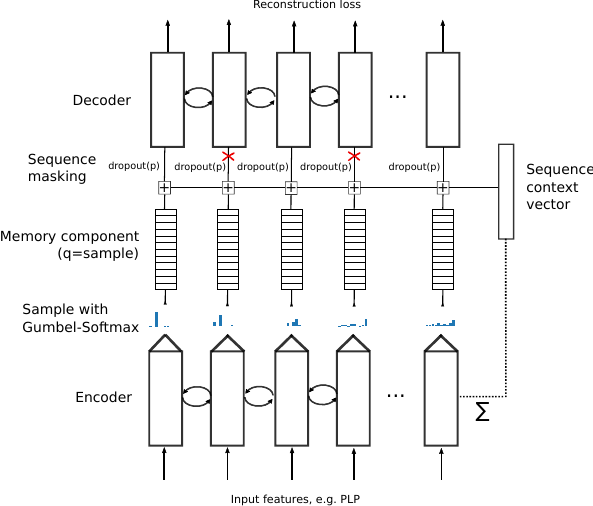}
  \caption{The Sparsespeech unsupervised acoustic model with Gumbel-Softmax.}
  \label{fig:model}
\end{figure}

The temperature parameter $\tau$ controls the amount of sparsity of the sample drawn from the distribution provided by the (unscaled) input logits. We illustrate this in Figure \ref{fig:gumbel_ex} with example samples drawn from the same distribution with varying $\tau$. Lower temperatures ($0.05$, $0.1$, $0.2$) tend to make the drawn samples sparse, approximating a one hot vector, while higher temperatures ($2.0$, $5.0$) increase denseness and approximate a uniform distribution. 

While training the network we use annealing, starting with a higher temperature and slowly decreasing it to a cutoff value below $0.5$, for example $\tau = 2 \rightarrow 0.1$. In the Sparsespeech model, the Gumbel-Softmax replaces the regular softmax with the sparsity constraint. Figure \ref{fig:model} illustrates the complete Sparsespeech model with the added Gumbel-Softmax.

\section{Setup}

We use the newest version of Sparsespeech\footnote{https://gitlab.com/milde/sparsespeech/}, Tensorflow 1.8 and Python 3.6.9. The relevant changes necessary for the categorical reparameterization have been added to the original model and repository.
For evaluation we use the supplied auxiliary scripts of the Libri-Light corpus\footnote{https://github.com/facebookresearch/libri-light}, with minor enhancements, such as the possibility to use Kullback-Leibler (KL) \cite{kullback1951information} as a distance function in the ABX evaluation.\footnote{These changes have been made available to the authors of Libri-Light as a pull request.} KL is a better metric to compare pseudo-spectograms such as the ones our model generates, while the default cosine distance function of the Libri-Light scripts are better suited for comparing embedding representations.

We use 4-layer stacked biLSTM decoder/encoders in our Sparsespeech models, with a width of 256 neurons for all experiments. Perceptual linear predictive (PLP) \cite{hermansky1990perceptual} input features are computed with Kaldi \cite{povey2011kaldi} using the standard settings of 13 dimensions and 100 frames per second on (downsampled if necessary) 16kHz audio. The sparsity constraint of Sparsespeech is disabled (sparsity weight set to 0) for all experiments with the new model, while the diversity constraint of original model is kept and the diversity weight set to 100. We keep the 2-stage training approach of the original model where the model is pre-trained without the memory component. For the second training stage, we use temperature annealing while training the network: the $\tau$ parameter for Gumbel-Softmax is set to 2.0 and then slowly decreases by multiplying with an annealing factor (0.9999) every x batches. A cut off parameter, 0.1 or 0.2 in most experiments is set after which the annealing scheme stops.

\section{Evaluation}

We evaluate on two proposed evaluation tasks in Libri-Light \cite{kahn2019libri}: completely unsupervised and semi-supervised with limited supervision on English audio book read speech. We currently focus on the 600h (small) and 6000h (medium) subsets of untranscribed speech to train our models. In the unsupervised evaluation, we measure ABX error rates \cite{schatz2013evaluating, schatz2016abx}. This provides an error rate that measures how well the trained unsupervised representation can differentiate between same/different tri-phones within and across speakers, for example "bit" vs. "bat". It is also agnostic to the representation and can be evaluated on pseudo-labels as well as dense representations. We use the dev sets to calibrate parameters and test the best performing models on the test set. The ABX error measure uses DTW to compare two segments of different length, we use symmetric Kullback-Leibler divergence as local comparison function. This is the recommended distance function for posteriorgram-like representations \cite{dunbar2017zero, dunbar2019zero}. In the semi-supervised setting, we first train a Sparsespeech model on the unannoated data from Libri-Light. We then follow \cite{kahn2019libri} and evaluate with a simple phoneme classifier with Connectionist Temporal Classification (CTC) loss \cite{graves2006connectionist} that is trained on the representation with 10h of limited-resource phone labels. 

\begin{table}[th]
  \caption{ABX error rates on features/posteriograms generated by our model for the Libri-Light dev set with different temperatures $\tau$.}
  \label{tab:ABX1}
  \centering
  \resizebox{\columnwidth}{!}{
  \begin{tabular}{llrrrr}
   \toprule
   \textbf{Model or features} &  \textbf{Temp.} & \multicolumn{2}{c}{\textbf{within speaker}} & \multicolumn{2}{c}{\textbf{across speaker}} \\
   \cmidrule{3-6}
   			& 	$\tau$		&  clean & other & clean & other \\
   	\midrule
 	PLP Features &      -        &	11.12 & 15.08 & 25.87 & 33.74 \\
S6000h-n42-$\tau2\rightarrow0.1$ &      0.2		  &  12.66 & 15.52 & 18.86 & 24.84 \\  
'' &      0.8        &  11.04 & 13.65 & 17.02 & 23.01 \\
'' &      1.0        &  10.66 & 13.25 & 16.34 & 22.55\\     
'' &      2.0        &  9.57 & 12.15 & 14.73 & 20.68 \\ 
'' &      \textbf{3.0}        & \textbf{9.51} & \textbf{12.15}  & \textbf{14.41}  & \textbf{20.25}  \\ 
'' &      5.0        & 10.48  & 12.94 & 15.28  & 20.87  \\ 
   
   \bottomrule
  \end{tabular}
  }
\end{table}

\begin{figure}[t]
  \centering
  \includegraphics[width=\linewidth]{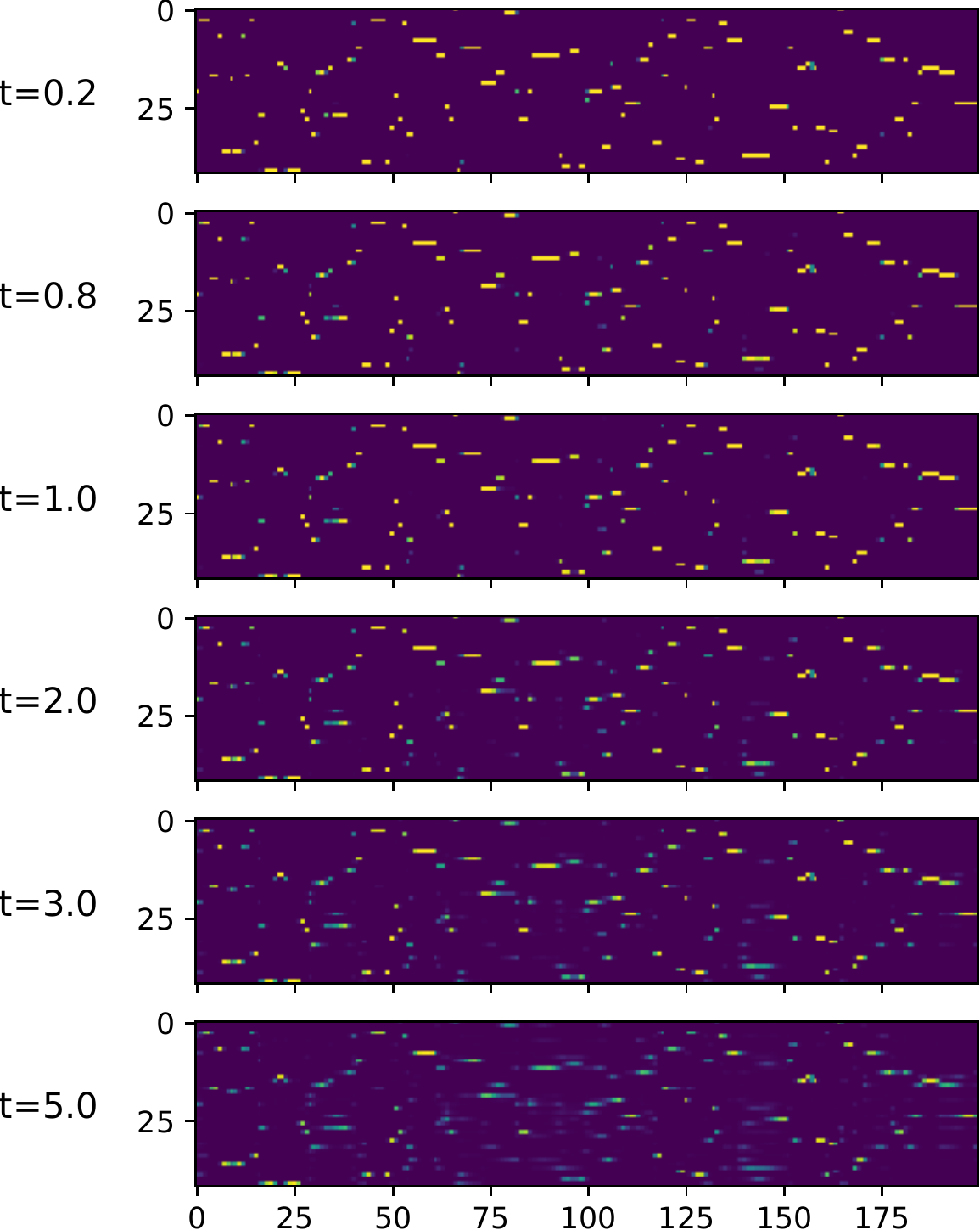}
  \caption{Example feature representations generated by the Sparsespeech model ''S6000h-n42-$\tau2\rightarrow0.1$'' with varying temperature.}
  \label{fig:sparseness}
\end{figure}

In Table \ref{tab:ABX1} we generate pseudo-posteriorgrams with different temperatures $\tau$ from the same model. This model has been trained on 6000h, with 42 components in the memory bank and output representation (n42) and a temperature annealing training scheme of $\tau2\rightarrow0.1$. The sparseness of the output can also be controlled with $\tau$ after training. The ABX error measure is sensitive to too sparse representations, as a different scaling with a higher $\tau$ significantly reduces the error measure. Temperatures 2.0 and 3.0 produced the lowest ABX error rates. 

\begin{table}[th]
  \caption{ABX error on features/posteriograms generated by our model for the Libri-Light dev set with different $n$ (components in the memory bank).}
  \label{tab:ABX2}
  \centering
  \resizebox{\columnwidth}{!}{
  \begin{tabular}{llrrrr}
   \toprule
   \textbf{Model or features} &  \textbf{Temp.} & \multicolumn{2}{c}{\textbf{within speaker}} & \multicolumn{2}{c}{\textbf{across speaker}} \\
   \cmidrule{3-6}
   			& 	$\tau$		&  clean & other & clean & other \\
   	\midrule
 	PLP Features (n=13) &      -        &	11.12 & 15.08 & 25.87 & 33.74 \\

	S600h-n20-sparsityloss-2.0     &     -     & 14.65 &  17.37 &  27.09  & 32.43   \\	
	
	S600h-n100-sparsityloss-2.0     &     -     & 14.03 &  16.63 &   24.80 & 29.67  \\	

   	\midrule

    S600h-n20-$\tau2\rightarrow0.5$ &      2.0		  & 11.56 & 13.75 & 21.18 & 26.66 \\    
    S600h-n42-$\tau2\rightarrow0.2$ &      3.0		  & 11.38 & 13.49  & \textbf{17.64} & \textbf{22.46} \\
    S600h-n100-$\tau2\rightarrow0.2$ &      2.0		  & 10.43 & 13.00 & 18.78  & 24.00 \\  
    S600h-n100-$\tau2\rightarrow0.2$ &      3.0		  & 10.43 & 13.00 & 18.86  & 24.08 \\  
   S600h-n128-$\tau2\rightarrow0.2$ &      3.0		  & \textbf{10.00} & \textbf{12.46} & 17.97  & 23.16 \\  
    S600h-n256-$\tau2\rightarrow0.2$ &      3.0		  & 11.41 & 14.02 & 22.73  & 27.55 \\   
   \bottomrule
  \end{tabular}
  }
\end{table}

In Table \ref{tab:ABX2} we mainly evaluate different $n$ in the memory component of Sparsespeech, trained on the Libri-Light small subset of 600h. Using n=100 or n=128 components produced good within and across speakers results, n=42 also performed well on the across speakers ABX error. All models have been trained for 3 epochs, with a training time ranging from 23.8h to 30.43h for the second stage of the Sparsespeech training (with the memory component) on a single Nvidia Titan XP GPU. We have experimented with different annealing schemes, but settled on $\tau = 2.0 \rightarrow 0.2$ for most experiments. We have also trained two Sparsespeech models with $n=20$ and $n=100$ using original sparsity loss training method without Gumbel-Softmax. This did not show good ABX error rates on the Libri-Light dev set, in fact only accross speaker ABX error improved over the PLP features baseline. The new models trained with Gumbel-Softmax show significant relative error rate improvements over the original Sparsespeech model, with nearly all tested representations better than PLP features in all tested settings.

\begin{table}[th]
  \caption{ABX error on features/posteriograms generated by our model for the Libri-Light test set. CPC results are taken from \cite{kahn2019libri}.}
  \label{tab:ABX_test}
  \centering
  \resizebox{\columnwidth}{!}{
  \begin{tabular}{llrrrr}
   \toprule
   \textbf{Model or features} &  \textbf{Temp.} & \multicolumn{2}{c}{\textbf{within speaker}} & \multicolumn{2}{c}{\textbf{across speaker}} \\
   \cmidrule{3-6}
   			& 	$\tau$		&  clean & other & clean & other \\
   	\midrule
 	PLP Features (n=13) &      -        &	10.46 &  14.69 & 23.78 & 34.15 \\

	S600h-n20-sparsityloss-2.0  & - & 13.98 & 16.95 & 24.80 & 32.66 \\
	
	S600h-n100-sparsityloss-2.0     & - & 14.12  & 16.97  & 22.86   & 30.53  \\

   	\midrule
 
    S600h-n20-$\tau2\rightarrow0.5$ &      2.0	   & 10.92 & 14.06  & 18.86  & 27.16 \\ 
 	 
    S600h-n128-$\tau2\rightarrow0.2$ &      3.0	   & 10.59 & 13.78 & 15.68  & 23.16 \\

	S6000h-n42-$\tau2\rightarrow0.1$ &      3.0	   & \textbf{9.33}  & \textbf{12.05}  & \textbf{13.53}  &  \textbf{20.60} \\

	\midrule
	
	CPC-600h (n=256)  &      -	   &  6.90 & 9.59 & 9.00  & 15.10 \\
	CPC-6000h (n=256) &      -	   &  6.22 & 8.55 & 8.05 & 13.81 \\
	CPC-60000h (n=256) &      -	   &  \textbf{5.83} & \textbf{8.14} & \textbf{7.56} & \textbf{13.42} \\

   \bottomrule
  \end{tabular}
  }
\end{table}

In Table \ref{tab:ABX_test} we compare ABX error rates on the Libri-Light test set. We use the best Sparsespeech models with 600h and 6000h as determined on the dev set. We compare against baseline PLP features, a baseline Sparsespeech model trained with the original sparsity loss method without Gumbel-Softmax and representations from Contrastive Predictive Coding (CPC) \cite{oord2018representation} as reported in \cite{kahn2019libri}. While the Sparsespeech models are trained on PLP features (n=13) as input, the CPC are trained on raw 16kHz speech in the time domain. Like on the dev set, there is a large reduction in error rates when training with Gumbel-Softmax. However, the dense embedding representations trained with the CPC model show lower error rates than the best Sparsespeech model that we trained on the 6000h medium subset.

\begin{table}[th]
  \caption{PER error for training a very simple phoneme recognizer with 10h of data on: PLP features, CPC model features or Sparsespeech model features.}
  \label{tab:PERerror}
  \centering
  \resizebox{\columnwidth}{!}{
  \begin{tabular}{llrrrr}
   \toprule
   \textbf{Model or features} &  \textbf{Temp.} & \multicolumn{2}{c}{\textbf{dev PER}} & \multicolumn{2}{c}{\textbf{test PER}} \\
   \cmidrule{3-6}
   			& 	$\tau$		&  clean & other & clean & other \\
   	\midrule
 	PLP Features (n=13) &      -        &	52.44 & 62.36 & 50.96  & 63.13 \\
	S600h-n100-sparsityloss-2.0    & - & 52.48 & 61.65 & 50.48 & 63.23 \\ 
   	\midrule
 	CPC-600h (n=256) &      -        & 40.21	 & 51.80  & 38.18  & 53.85 \\
	CPC-6000h (n=256) &      -        & 34.40	 & 47.60 & 34.44 & 49.40 \\
    CPC-60000h (n=256) &      -         &	31.16 & 46.67 & 32.67 & 48.93  \\

   	\midrule
	S600h-n100-$\tau2\rightarrow0.2$  &      3.0        & 50.39	 & 59.82 & 48.29 & 61.75 \\
	S600h-n128-$\tau2\rightarrow0.2$  &      3.0        & 50.56 & 60.20 & 48.05 & 61.69 \\
    S6000h-n42-$\tau2\rightarrow0.1$  &      3.0        & 47.77	 & 57.77 & 46.61  & 59.61  \\
   \bottomrule
  \end{tabular}
  }
\end{table}

In Table \ref{tab:PERerror} we compare the representations of our models in terms of how well a simple phoneme recognizer can classify phonemes with it as input. The phoneme recognizers are trained on 10h of representations with phone labels. They are trained without explicit alignments using the CTC loss. Using only a linear 1D-convolution on posteriorgram-like representations as in \cite{kahn2019libri} proved to be challenging, as the most frequent emission symbol per timestep with the CTC loss is the blank label. Adding a simple 1-layer LSTM makes sure that the network can learn when to emit a label other than the blank label and also keep track of context. The 1D convolution has a kernel size of 8 (default in the Libri-Light evaluation script) and the number of output channels of the convolution is set to match the number of phones in the transcription plus the blank label (45). The 1-Layer LSTM has a fixed hidden size of 100.

A simple decoder with beam search generates the hypothesized phone sequence. Phoneme error rate (PER) is then computed by comparing the sequence to the Libri-Light transcriptions on the dev and test set (these sets are the same as the ones in Librispeech \cite{panayotovlibrispeech}). With the original Sparsespeech model we were getting similar PER results as the PLP baseline, but with the improved Sparsespeech model the phoneme recognizer can improve PER by 8.5\% relative over the PLP baseline on the Libri-Light test-clean and by 5.6\% relative on test-other.

\section{Conclusion}

Using Gumbel-Softmax in the Sparsespeech model is an effective improvement. It yields a relative reduction of 23.9\% in ABX error rates on the test set (with clean speech) accross speakers compared to the original model in \cite{milde2019sparsespeech} on $n=20$. The dimensionality of the learned representations of the new Sparsespeech model can now also be scaled up to representations with bigger $n$, while also improving ABX error rates. So far $n=100$ and $n=128$ yielded the best results for within speaker ABX when trained on 600h of untranscribed speech. Representations with bigger $n$ also show a relative improvement of over 31.4\% on ABX error rates across speakers on the clean test set compared to the original model. A first result on training on the 6000h medium subset of the Libri-Light corpus further improved error rates and shows that the model is scaling. Currently, tuning the temperature parameter after a Sparsespeech model has been trained seems to be important to reduce ABX error rates, but higher temperatures when generating Sparsespeech features such as 2.0 and 3.0 seem to work well across models with different hyperparameters.

PER error rates also show an 8.5\% improvement over a PLP baseline with the new model when a simple phoneme recognizer is trained on the representations. The generated representations from the new model are still relatively compact and sparse (see also Figure \ref{fig:sparseness}) with better phoneme discriminability as measured by ABX and PER than PLP features. However when we compare ABX and PER error to unsupervised dense embedding representations such as the ones generated by CPC (n=256), there still a relatively large gap in error rates on the Libri-Light test set.

One difference is the type of input features; CPC uses raw wavforms in the time domain while we have used PLP features. We also plan to try out end-to-end learning on raw wavforms, as this could show how much of the performance gap can be attributed to this difference. CPC representations could potentially also be used as input features to the Sparsespeech model or the models could be combined. We are currently also working on scaling the Sparsespeech model to the full 60 thousand hours of speech data available in Libri-Light. Based on the training time on the smaller sets, we would expect an estimated training time of roughly 10 weeks on a single GPU with Sparsespeech for the full Libri-Light dataset. 

Another major difference is strucural in the type of the generated representations. There might be a trade-off in ABX error rates between low-bitrate sparse representations and higher bitrate dense representations. The results from last years Zero Resource Challenge \cite{dunbar2019zero} support this hypothesis with systems with higher ABX errors having lower bit rate representations. The organizers concluded that ''this suggests that discretizing learned speech embeddings well is hard''. 

The pseudo-posteriorgrams that our Sparsespeech model can generate have the advantage over embeddings that they can directly be interpreted as a (soft) clustering of phoneme-like units. They can also be easily discretized and translated to symbolic pseudo transcriptions, where the ABX discriminability is still largely preserved (see \cite{milde2019sparsespeech}).

\bibliographystyle{IEEEtran}
\bibliography{mybib}

\end{document}